\documentclass[titlepage,12pt]{utarticle}

\usepackage{graphics}
\usepackage{latexsym,amsmath,amssymb,amscd,cite}
\usepackage{epsf}
\usepackage[all]{xy}
\usepackage{hyperref}

\DeclareFontFamily{U}{rsf}{}
\DeclareFontShape{U}{rsf}{m}{n}{<5> <6> rsfs5 <7> <8> <9> rsfs7 <10-> rsfs10}{}
\DeclareMathAlphabet\Scr{U}{rsf}{m}{n}

\newcommand{\al}{\alpha}
\newcommand{\be}{\beta}

\newcommand{\de}{\delta}

\newcommand{\et}{\eta}

\newcommand{\f}{\phi}
\newcommand{\vf}{\varphi}
\newcommand{\g}{\chi}

\newcommand{\m}{\mu}

\newcommand{\p}{\pi}

\newcommand{\rh}{\rho}

\newcommand{\s}{\sigma}

\newcommand{\ta}{\tau}

\newcommand{\om}{\omega}

\newcommand{\y}{\psi}

\newcommand{\De}{\Delta}

\newcommand{\Si}{\Sigma}

\newcommand{\Y}{\Psi}

\newcommand {\cA} {{\cal A}}

\newcommand {\cC} {{\cal C}}

\newcommand {\cF} {{\cal F}}

\newcommand {\cH} {{\cal H}}

\newcommand {\cM} {{\cal M}}
\newcommand {\cN} {{\cal N}}

\newcommand {\cW} {{\cal W}}

\newcommand {\bbC} {\mathbb{C}}

\newcommand {\bbZ} {\mathbb{Z}}

\newcommand{\bG}{{\bar G}}
\newcommand{\bH}{{\bar H}}
\newcommand{\bJ}{{\bar J}}

\newcommand{\bQ}{{\bar Q}}
\newcommand{\bW}{{\bar W}}

\newcommand{\ba}{{\bar a}}

\newcommand{\bt}{{\bar t}}

\newcommand{\bz}{{\bar z}}

\newcommand{\hc}{{\hat c}}
\newcommand{\hg}{{\hat g}}

\newcommand{\Tr}{{\rm Tr}}

\newcommand{\td}{\tilde}

\newcommand{\unit}{{1\!\!\!\!1\,}}

\newcommand{\dl} {\partial}

\newcommand{\der}[1]{\frac{\partial}{\partial {#1}}}

\newcommand {\bra}{\bigl\langle}
\newcommand {\ket}{\bigr\rangle}

\newcommand{\braarg}[1]{\bra\,#1\,\big|}
\newcommand{\ketarg}[1]{\big|\,#1\,\ket}

\newcommand{\rarrow}{\rightarrow}

\long\def\beq #1 \eeq{\begin{equation}#1\end{equation}}
\long\def\bea #1 \eea{\begin{eqnarray}#1\end{eqnarray}}
\long\def\beann #1 \eeann{\begin{eqnarray*}#1\end{eqnarray*}}

\long\def\ba #1 \ea{\begin{array}#1\end{array}}

\newcommand{\bit}{\begin{itemize}}
\newcommand{\eit}{\end{itemize}}

\newcommand{\ben}{\begin{enumerate}}
\newcommand{\een}{\end{enumerate}}

\long\def\bec #1 \eec{\begin{center}#1\end{center}}

\newtheorem{theorem}{Theorem}
\newtheorem{define}{Definition}

\long\def\bth #1 \eth{\begin{theorem}#1~\end{theorem}}
\long\def\bdef #1 \edef{\begin{define}#1~\end{define}}
\long\def\bpr #1 \epr{~\\ \emph{Proof:}~ #1~$\Box$}

\long\def\del#1\enddel{}
\long\def\new#1\endnew{{\bf #1}}

\def\ll{\label}
\def\nn{\nonumber{}}

\newcommand{\tfor}{{\qquad \mathrm{for} \quad}}

\newcommand{\Acat}{{\cA^q_\infty}}

\newcommand{\lcder}{\stackrel{\rightharpoonup}{\partial}}

\newcommand{\rcder}{\stackrel{\leftharpoonup}{\partial}}

\newcommand{\ie}{\textit{i.e.}}

\begin{document}        

\preprint{
  DESY-06-013\\
  {\tt hep-th/0602018}\\
}
\title{Quantum $\cA_\infty$-Structures\\[5pt] 
       for Open-Closed Topological Strings}
\author{Manfred Herbst} 
    \oneaddress{
     DESY Theory Group\\
     Notkestra\ss e 85\\
     22603 Hamburg\\
     Germany\\
     \email{Manfred.Herbst@desy.de}\\
    }

\Abstract{
  We study factorizations of topological string amplitudes on higher genus
  Riemann surfaces with multiple boundary components and find quantum
  $\cA_\infty$-relations, which are the higher genus analog of the (classical)
  $\cA_\infty$-relations on the disk. For topological strings with $\hc=3$ the
  quantum $\cA_\infty$-relations are trivially satisfied on a single D-brane,
  whereas in a multiple D-brane configuration they may be used to compute open
  higher genus amplitudes recursively from disk amplitudes.
  This can be helpful in open Gromov--Witten theory in order to
  determine open string higher genus instanton corrections.\\
  Finally, we find that the quantum $\cA_\infty$-structure cannot quite
  be recast into a quantum master equation on the open string moduli space.
}
\date{}

\maketitle
\tableofcontents

\newpage

\section{Introduction and summary}
\ll{sec:intro}

The perturbative topological closed string is by now quite well-studied. The
holomorphic anomaly equations \cite{BCOV93KS} suggest an
interpretation of the topological string partition function as a wave
function \cite{W93wave,V04wave}, and current research is concentrating 
on understanding this wave function in terms of a non-perturbative completion
of the topological string.

The open topological string is far from such an understanding --- it is not
even well-studied perturbatively. This work is a first step in this direction.
We do not yet try to consider the analog of holomorphic anomalies for open
strings; these are notoriously difficult to handle (cf. the remarks in
\cite{BCOV93KS}). Instead we concentrate on a more 
fundamental difference between open and closed topological strings.

Deformations of closed string moduli are not obstructed, \ie, they are
not subject to some potential. This is reflected in the fact that the
associated on-shell%
\footnote{By \emph{on-shell} we mean the minimal $L_\infty$ structure on the
  space of topological observables, \ie, the cohomology classes of the
  BRST operator. 
}
$L_\infty$ structure vanishes \cite{KS05OCHA}, which
is the main reason for why $L_\infty$ structures played a minor
r\^ole in the physical literature on topological closed strings.

The picture is quite different in open topological string
theory (at tree-level). The open string moduli $s^a$ are obstructed because
they are lifted by an effective superpotential
\cite{Laz01Superpot}, 
\[  \cW_{eff} = \sum_{m=1}^\infty 
    \frac 1m \cF_{a_1 \ldots a_m} s^{a_1}\ldots s^{a_m}
\]
which is due to a non-vanishing on-shell $\cA_\infty$
structure for the disk amplitudes $\cF_{a_1 \ldots a_m}$
\cite{HLL04Ainfty}. Therefore, in contrast to closed topological
strings homotopy algebras play a more important r\^ole in open
topological string theory.

In \cite{HLL04Ainfty} the $\cA_\infty$-structure was derived from
the bubbling of disks, which was induced by the insertion of the topological
BRST operator. In the present work we extend this analysis to topological
string amplitudes $\cF^{g,b}_{A_1|\ldots|A_b}(t)$ on genus $g$ Riemann
surfaces with $b$ boundary components. Here, $A_i$, for $i=1,\ldots,b$, is a
collective index for the cyclically ordered topological observables, which are
inserted at the $i^{th}$ boundary circle. $t$ is a closed string modulus. It
turns out to be convenient to 
resume these amplitudes in the all-genus topological string amplitudes,
\[ \cF^{b}_{A_1|\ldots|A_b}(g_s,t) = 
   \sum_{g=0}^\infty g_s^{2g+b-2} \cF^{g,b}_{A_1|\ldots|A_b}(t) \ ,
\]
where we introduced the topological string coupling constant $g_s$.
The main result of this work then states that \emph{the all-genus topological
string amplitudes $\cF^{g,b}_{A_1|\ldots|A_b}(g_s,t)$ satisfy the (cyclic)
quantum $\cA_\infty$-relations (\ref{qAinftysum})}.

At first sight these relations look quite complicated, but they bear some
interesting features:
\ben
  \item There are \emph{no closed string factorization channels} involved in
  these relations. This is essentially due to the fact that
  degenerations in the closed string channel correspond to (real)
  codimension $2$ boundaries of the moduli space of the
  Riemann surface, whereas the insertion of the BRST operator
  'maps' only to the (real) codimension $1$ boundary. This will be
  explained in detail in the present work.
  The quantum $\cA_\infty$-structure is, therefore, simpler than the
  homotopy algebra structures that appear in Zwiebach's open-closed
  string field theory \cite{Zwie97OCSFT,KS05OCHA}.

  \item As it was pointed out in \cite{AK04Superpot}, in the context of the
  topological B-model, the (classical) $\cA_\infty$-relations are
  \emph{trivially satisfied on a single D-brane}. We find that this is 
  true for the full quantum $\cA_\infty$-relations in models with
  central charge $\hat c = 3$. This fact relies on
  arguments involving the $U(1)$ R-charge of the topological observables.

  \item On the other hand, in a situation with \emph{multiple D-branes} the
  quantum $\cA_\infty$-relations give rise to a sequence of linear systems for
  the amplitudes $\cF^{g,b}_{A_1|\ldots|A_b}(t)$, which can be solved
  \emph{recursively} starting from disk amplitudes. 
%%   However, the linear system
%%   may be underdetermined, so that additional physical information must be
%%   provided in order to get a unique solution. 
  As an example, we find solutions to the linear system in the context
  of open string instanton counting on the elliptic curve in the companion
  paper \cite{HLN06torus}.
\een

For the derivation of the quantum $\cA_\infty$-relations we make
the technical assumption that every boundary component of the Riemann surface
carries at least one observable. This is important, for otherwise there
appear non-stable configurations, which correspond to non-compact
directions in the moduli space of the
Riemann surface. More specifically, these come from closed string
factorization channels where a 'bare' boundary component, \ie, without 
operator insertions, bubbles off from the rest of the Riemann
surface. The appearance of a non-stable configuration indicates
ambiguities in view of divergences of topological string amplitudes
(cf. \cite{HLN06torus}).

As it is familiar from string field theory \cite{Zwie97OCSFT,Kaj01SFT} and
recently from open topological string theory \cite{Laz05NCgeom}, homotopy
algebras have a dual description in terms of a master equation on a dual
supermanifold (in the present context, the open string moduli
space). We briefly comment on this relation and find 
that we have to formally include the non-stable configurations, that we just
alluded to, in order to recast the quantum $\cA_\infty$-structure into the
modified quantum master equation (\ref{QME}). We want to stress that on
the way of deriving this relation some information on the
topological string amplitudes is lost so that the quantum
$\cA_\infty$-structure is not faithfully mapped to the
modified quantum master equation.

%% In order to make contact with type II superstring compactifications with
%% D-branes we collect some results from the literature 
%% \cite{AGNT93TST,BCOV93KS,Vafa00largeN,OV03Cdef,OV03gravCdef} 
%% and give a proposal for what kind of F-terms the open higher genus
%% topological 
%% string amplitudes give rise to in $N=1$ supergravity.

Before giving an outline of the paper we list some concrete models where the
quantum $\cA_\infty$-relations may be applied.
\emph{(i)} Consider the topological $B$-model on a Calabi--Yau
$3$-fold $\cM$ in the presence of $B$-type boundary conditions. The latter
correspond to holomorphic vector bundles over holomorphic submanifolds or,
more generally, to objects in the derived category of coherent sheaves on
$\cM$ \cite{Kon94HMS,Douglas00DCategory}. At the classical level the
$\cA_\infty$-structure was computed for some explicit D-brane configurations
on non-compact Calabi--Yau manifolds in \cite{AK04Superpot}. \emph{(ii)}
The mirror dual of such a model is the topological $A$-model
on $\cM$ with $A$-type boundary conditions describing
Lagrangian submanifolds and the coisotropic D-branes of
\cite{KO01coiso,KO03coisoLect}. These are objects in Fukaya's $\cA_\infty$
category \cite{Fukaya93AinftyCat}. Correlation functions in the $A$-model
receive world-sheet instanton corrections, thus relating it to
enumerative geometry of rational curves in $\cM$ or 
holomorphic disks spanned between Lagrangian submanifolds. In this context the 
quantum $\cA_\infty$-relations could provide a means of recursively computing
higher genus open Gromov--Witten invariants from disk invariants. 
\emph{(iii)} A last example are topologically twisted Landau--Ginzburg
(LG) orbifolds with $B$-type boundary conditions. The D-branes correspond to
equivariant matrix factorizations of the LG superpotential
\cite{%KontsevichLG,Orlov03LG,BHLS03LG,KL03LG,
ADD04Fractional,HW04Fterm}.

The paper is organized as follows: In section \ref{sec:defampl}
open-closed topological string amplitudes are defined, and basic
symmetry properties as well as selection rules are reviewed. The main
result, the quantum $\cA_\infty$-relations for the amplitudes
$\cF^{b}_{A_1|\ldots|A_b}(g_s,t)$, is stated and proven in section
\ref{sec:qAinfty}. We proceed with the discussion of some features 
of our result in section \ref{sec:comments} and close with the
derivation of the modified quantum master equation on the open string
moduli space in section \ref{sec:QME}.

\section{Open-closed topological string amplitudes}
\ll{sec:defampl}

Let us start with reviewing the general setup of open-closed topological
string theory. This will give us also some room to introduce notations. 
By definition a topological string theory is a 2d topological
conformal field theory (TCFT) coupled to gravity.

So let us have a look at the TCFT first. The most important
relation in the topological operator algebra \cite{DVV90DVV} can be
stated as the fact that the stress-energy tensor $T_{zz}$ is
BRST exact, \ie,%
\footnote{Subsequently, we use $[.,.]$ as a graded commutator.}
\beq
  \ll{OPA}
  T_{zz}(z) = [Q, G_{zz}(z)] \ .
\eeq
Here, $Q$ is the BRST operator and $G_{zz}$ is the fermionic current
of the operator algebra. The $U(1)$ R-current $J_z(z)$ does not
interest us for the moment, but will play an important role
in subsequent sections.

Since we want to consider models on general oriented, bordered Riemann
surfaces $\Si$ we have to specify boundary conditions on the
currents. The only choice for picking these boundary
conditions comes from the $U(1)$ automorphism of the topological 
operator algebra, which acts as a phase factor $e^{i\p\vf}$ on the
fermionic currents $Q(z)$ and $G_{zz}(z)$. However, a single-valued
correlation function requires that the difference of the phases
between two boundary conditions is integral, $\De \vf \in \bbZ$
(cf. \cite{RS97GPbranes} in the context of superconformal
algebras). The overall 
phase is unphysical and can be set to zero, so that the currents of
the topological operator algebra satisfy the simple relations
\beq
  \nn
  \begin{array}{l@{~=~}l}
    W^{bos} & \bW^{bos} \\
    W^{ferm} & (-)^s \bW^{ferm}
  \end{array}
  \quad\textrm{on}\quad \dl \Si \ ,
\eeq
for integral $s$.

The observables of an open-closed string theory are the cohomology
classes of the BRST operator $Q$. \emph{Bulk observables} $\f_i$ are
in one-to-one correspondence with  states in the closed string Hilbert
space $\cH_{cl}$, in short we write $\f_i \in \cH_{cl}$. Whereas
\emph{boundary observables} $\y^{\al\be}_a$ correspond to a states in the
open string Hilbert space $\cH^{\al\be}_{op}$. The upper indices
denote the boundary conditions, \ie, the topological 
D-branes, on either side of the field. The cyclical order of the
boundary fields ensures that the boundary labels $\al, \be, \ldots$, once
determined, always match, so that we restrain ourselves from writing
the boundary labels explicitly in the following. In a topological
\emph{conformal} field theory it is important to choose a particular
representative of the BRST cohomology class by requiring \cite{DVV90DVV}
\beq
  \ll{gauge}
  [G_0,\f_i] =[\bG_0,\f_i] = [(G_0+(-)^s\bG_0),\y^{\al\be}_a]=0 \ ,
\eeq which implies that the topological
observables have conformal weight zero.

In view of relation (\ref{OPA}) we can define topological descendents 
\cite{Witten88TFT} associated with bulk and boundary observables: the
bulk $2$-form descendent $\f^{(2)}_i$ and the boundary $1$-form descendent
$\y^{(1)}_a$ are particularly important for us and satisfy the relations
\beq
  \ll{descendents}
  [Q, \f^{(2)}_i] = d \f^{(1)}_i \quad\textrm{resp.}\quad
  [Q, \y^{(1)}_a] = d \y_a \ .
\eeq

The coupling of the TCFT to 2d gravity is similar to
the bosonic string and we can inherit the procedure to define
amplitudes on bordered higher genus Riemann surfaces.%
\footnote{We are not considering gravitational descendents here
  \cite{Witten89TopGrav}!}
For that we introduce the quantities 
\beq
  \ll{Beltrami}
  G_{m_i} := \int_{\Si} d^2 z~ 
             (G_{zz}\m_i^z{}_\bz + 
              G_{\bz \bz} \m_i^\bz{}_z) \ ,
\eeq
which we have to insert in the path integral to account for zero modes
related to the complex structure moduli space of the Riemann surface
$\Si$. In (\ref{Beltrami}) $\m_i^z{}_\bz$ denotes the Beltrami differential,
\[ \m_i^z{}_\bz :=  \frac12 g^{z\bz} \der{m_i} g_{\bz \bz} \ ,
\]
corresponding to the complex structure modulus $m_i$.
Using (\ref{OPA}) and the definition of the stress-energy tensor, we
deduce the relation
\beq
  \ll{Gdesc}
  \bra\ldots [Q,G_{m_i}] \ldots \ket = \der{m_i} \bra\ldots\ket \ .
\eeq
% Note the reminiscence of this equation with \eqref{descendents}; in both
% cases $Q$ gives rise to a derivative. 

\subsection{Definition of the amplitudes}

Having set up the basics we can start defining 
the topological string amplitude on oriented Riemann
surfaces $\Si_{g,b}$ of genus $g$ with $b$ boundary components.%
\footnote{ In the following we are not interested in the holomorphic
  anomaly of the higher genus amplitudes \cite{BCOV93KS}, which means
  that we \emph{fix} a particular background $\bt=\bt_0$ where the
  theory is well-defined and do not consider changes thereof. We will
  not indicate this background value subsequently and we will work in
  flat coordinates.}

A measure on the moduli space of punctured bordered Riemann surfaces
is defined consistently only for \emph{stable} configurations, that
is, if
\beq 
  \ll{stable}
  2g+b-2+ 2n + \sum_{i=1}^b m_i \geq 0 \ ,
\eeq
where $n$ is the number of bulk observables and $m_i$ is the number of
boundary observables inserted on the $i^{th}$ boundary component. 

Riemann surfaces with Euler character $\g_{g,b}=2-2g-b \leq -1$ do not
possess conformal Killing vector fields, so that we are not forced to
gauge the corresponding symmetry. Following the prescription in
bosonic string theory the amplitudes are defined as
\beq
  \ll{highergb}
    \cF^{g,b}_{A_1|\ldots|A_b}(t) := \int d^{3|\g_{g,b}|}m~
    \bra P\!\!\int\!\! 
    \y_{a[1]_1}^{(1)}\ldots \int\!\! \y_{a[1]_{m_1}}^{(1)}\big|
    \!\int\!\! \y_{a[2]_1}^{(1)}\ldots \big| \ldots \big|
    e^{t^i \int\f^{(2)}_i}
    \prod_{i=1}^{3|\g_{g,b}|} G_{m_i} \ket_{\Si_{g,b}} \ ,
\eeq
where we introduced the collective index
$A_i = a[i]_1\ldots a[i]_{m_i}$. $P$ means that the boundary
descendents are integrated in cyclical order; subsequently we will
refrain from writing $P$ explicitly. The integration over the complex
structure moduli space of the punctured Riemann surface will be
defined more carefully later in section \ref{sec:modulispace}. 

The amplitudes (\ref{highergb}) can be understood  
as generating function for all amplitudes with arbitrary number of
bulk insertions. We will sometimes refer to the amplitudes
(\ref{highergb}) as deformed amplitudes as compared to undeformed ones
where the closed string moduli $t^i$ are turn off, \ie, $t^i=0$. 
As already emphasized in \cite{HLL04Ainfty} it is \emph{not}
possible to subsume the boundary observables as deformations because
of the boundary condition labels and the cyclic ordering.

There are four Riemann surfaces which do have conformal Killing
vectors and need special treatment. Let us start with the simplest
one, the Riemann sphere $\Si_{0,0}=S^2$. The global conformal group is
$SL(2,\bbC)$, which can be gauged by fixing three bulk
observables. The sphere three-point function defines the prepotential 
$\cF=\cF^{0,0}$ in the well-known way:
\beq
  \ll{sphere}
  \dl_i \dl_j \dl_k \cF^{0,0}(t) := 
  \bra \f_i~ \f_j~ \f_k~  e^{t^i \int\f^{(2)}_i} \ket_{\Si_{0,0}} \ .
\eeq
On the torus $\Si_{1,0}$ the conformal Killing vectors correspond to
translations and the complex structure moduli space $\cM_{1,0}$ is the
upper half complex plane mod $PSL(2,\bbZ)$. The amplitude reads:
\beq
  \ll{torus}
  \dl_i \cF^{1,0}(t) := \int d\tau d\bar\tau
                  \bra \f_i ~  e^{t^i \int\f^{(2)}_i}~
		  G_\tau~ G_{\bar\tau} \ket_{\Si_{1,0}} \ .
\eeq

For the bordered Riemann surfaces with conformal Killing
vectors the situation is a bit more complex for the reason that we can
use either bulk or boundary observables to fix the global conformal
symmetries. On the disk $\Si_{0,1}$ we need to fix three real
positions, so that we can have two types of amplitudes (which are,
however, related by Ward identities \cite{HLL04Ainfty}):%
\footnote{We anticipate the suspended grading $\td a_i$ from
  (\ref{suspgrade}), which is the fermion number of the descendent
  $\y_{a_i}^{(1)}$. Note the overall sign change in (\ref{disk3}) as
  compared to \cite{HLL04Ainfty}.
}
\beq
  \ll{disk3}
  \cF^{0,1}_{a_1 \ldots a_m}(t) := -(-1)^{\td a_2 + \ldots + \td a_{m-1}}
  \bra \y_{a_1} \y_{a_2} P \int \y_{a_3}^{(1)} \ldots 
  \int \y_{a_{m-1}}^{(1)} \y_{a_m} ~  e^{t^i \int\f^{(2)}_i}\ket_{\Si_{0,1}} \ ,
\eeq
for $m \geq 3$, and
\bea
  \ll{disk11}
  && \dl_i \cF^{0,1}_a(t) := 
     \bra \f_i \y_a ~  e^{t^i \int\f^{(2)}_i}\ket_{\Si_{0,1}} \ ,\\
  && \dl_i \cF^{0,1}_{ab}(t) := 
     \bra \f_i \y_a \int \y_b^{(1)}~  e^{t^i \int\f^{(2)}_i}\ket_{\Si_{0,1}}
  \nn \ .
\eea
On the annulus $\Si_{0,2}$ the rotation symmetry can be fixed by a
boundary observable or a bulk descendent $\f^{(1)}_i$ integrated along a
contour from one boundary to the other, resulting in:
\beq
  \ll{annulus1}
  \cF^{0,2}_{A_1 | A_2}(t) := 
  \int_0^\infty dL 
  \bra \y_{a[1]_1}  \!\!\int\!\! \y_{a[1]_2}^{(1)} \ldots 
  \!\!\int\!\! \y_{a[1]_{m_1}}^{(1)} \big|
  \!\!\int\!\! \y_{a[2]_1}^{(1)} \ldots 
  \!\!\int\!\! \y_{a[2]_{m_2}}^{(1)}\big|~  
  e^{t^i \int\f^{(2)}_i} G_L \ket_{\Si_{0,2}} \ ,
\eeq
for $m_1 \geq 1$. For annulus amplitudes without any boundary
insertions (cf. \cite{BCOV93KS}) we define:
\beq
  \ll{annulus2}
  \dl_i \cF^{0,2}_{.|.}(t) := \int_0^\infty dL \bra \int_\cC \f_i^{(1)}~ 
                 e^{t^i \int\f^{(2)}_i}~G_L \ket_{\Si_{0,2}} \ ,
\eeq
where the contour $\cC$ runs from one boundary component to the other.
The integration to $\cF^{g,b}_{\ldots}$ in
(\ref{sphere}), (\ref{torus}) as well as
(\ref{disk11}), (\ref{annulus2}) is well-defined 
through conformal Ward identities \cite{DVV90DVV,HLL04Ainfty}. In the
following we will assume that the integration constants in
(\ref{disk11}) are zero, \ie, 
$\cF^{0,1}_a\big|_{t=0} = \cF^{0,1}_{ab}\big|_{t=0} = 0$.

For later convenience we introduce the all-genus topological string
amplitudes
\beq
  \ll{allgb}
  \cF^{b}_{A_1|\ldots |A_b}(g_s;t) := 
    \sum_{g=0}^\infty g_s^{2g+b-2}~ \cF^{g,b}_{A_1|\ldots |A_b}(t) 
    \tfor b \geq 1 \ .
\eeq
Note that for $b=0$ the analogous definition gives the topological
closed string free energy
$\cF(g_s;t) = \sum_{g=0}^\infty g_s^{2g-2}~ \cF^{g,0}(t)$ \cite{BCOV93KS}.

%% As shown in \cite{BCOV93KS} (also \cite{}) the topological amplitudes
%% $\cF^{g,0}$ satisfy holomorphic anomaly equations for $\hc =3$, which
%% could most concisely be written in terms of the free energy
%% \beq
%%   \ll{freeenergy}
%%   \cF^{\bul,0}(t,\bt) := \sum_{g=0}^\infty g_s^{2g-2}~ \cF^{g,0}(t,\bt) \ .
%% \eeq
%% or better:
%% \beq
%%   \ll{genfunction}
%%   W^{\bul,0}(t,\bt;x) := \cF^{\bul,0}(t\!+\!x,\bt) =
%%   \sum_{g=0}^\infty g_s^{2g-2}~ 
%%   \sum_{n=0}^\infty \frac{1}{n!} ~ 
%%   x^{i_1} \ldots x^{i_n} D_{i_1}\ldots D_{i_n} \cF^{g,0}(t,\bt) \ .
%% \eeq

%% (how are holom. anomaly equations derived -> remark that we use a
%% similar method of contour deformation here.)

\subsection{Topological twist and charge selection rules}

Suppose we started with an $\cN=(2,2)$ superconformal algebra (broken
by boundary conditions to $\cN=2$), which is part of a superstring
compactification. We will leave the central charge $c=3\hc$ arbitrary
for the moment. The topological twist by 
$T \rarrow T + \frac12 \dl J$ to the associated topological
algebra is implemented by coupling the spin connection
$\om=\dl\ln(\sqrt{g})$ to the $U(1)$ current $J$ in the action
\cite{Witten88TFT,Witten88TopSigma,BCOV93KS}, \ie, 
\[ \frac{1}{8\p}\int_{\Si_{g,b}} d^2z (\om \bJ + \bar\om J) \ .
\]
Using the bosonization $J = i\sqrt{\hc} \dl H$ and taking into account
that we have boundaries we get
\beq
  \ll{bgcharge}
  - \frac{i\sqrt{\hc}}{8\p}\int_{\Si_{g,b}} d^2z \sqrt{g}R^{(2)} H
  + c.c. - \frac{i\sqrt{\hc}}{4\p}\int_{\dl\Si_{g,b}} d\ta k H + c.c.\ ,
\eeq
where $R^{(2)}$ is the world sheet curvature and $k$ is the geodesic
curvature along the boundary. If we deform the world sheet metric such
that the curvature localizes at $|\g_{g,b}|$ points the twisting term
(\ref{bgcharge}) gives rise to $\g_{g,b}$ insertions of the spectral
flow operator $e^{-\frac{i}{2}\sqrt{\hc}(H-\bH)}$ in the
superconformal correlator, which, in total, carry the background
$U(1)$ charge $-\hc \g_{g,b}$.

In view of these considerations the $U(1)$ charges $Q_\al$ of all
operator insertions have to satisfy the condition
\[
  \sum_{\al=1}^{\#_{insert}} Q_\al = \hc \g_{g,b} \ .
\]
Let us consider an arbitrary Riemann surface $\Si_{g,b}$ with 
$\g_{g,b} < 0$. From the discussion in the previous section we know
that we have to insert $-3\g_{g,b}$ Beltrami differentials coupled to
$G_{zz}$, which have charge $Q_G = -1$.
Furthermore, every bulk descendent $\f^{(2)}_i$ carries
$Q_i=q_i - 2$, where $q_i$ is the charge of the associated topological
observable $\f_i$, and every boundary descendent $\y^{(1)}_a$ carries charge
$Q_a = q_a - 1$. The charge selection rule becomes
\beq
  \ll{chargeselect}
  \sum_{i=1}^n q_i + 
  \sum_{l=1}^b \sum_{a_l=1}^{m_l} q_{a_l} = 
  2n + m + (\hc-3) \g_{g,b} \ ,
\eeq
where $n$ and $m = \sum_{l=1}^b m_i$ are the total number of bulk
resp. boundary observables. Performing a case-by-case study it is easy
to show that  this formula extends to Riemann surfaces with $\g_{g,b}
\geq 0$, \ie, the sphere, the disk, the annulus and the torus.

\subsection{Background fermion number -  (suspended) $\bbZ_2$-grading}

The bulk as well as the boundary fields of a general topological
string theory carry a $\bbZ_2$-grading associated to the fermion
number $F$. Just as for the $U(1)$ charge we have to cancel a background
fermion number $\om \in \bbZ_2$, which accounts for the insertion of
fermionic zero modes in the path integral on the disk. In the context
of matrix factorizations in Landau--Ginzburg models this is manifest
in the Kapustin--Li formula \cite{KL03LG,HL04Local}, whereas in the
topological A- and B-model the background charge can be read off from
the inner product for the Chern--Simons resp. holomorphic
Chern--Simons action \cite{W92CS}.

A generalization of the $\bbZ_2$-selection rule to arbitrary Riemann
surfaces can readily be seen from a factorization argument within 2d
topological field theory \cite{MooreSegal,Laz00TFT}; every boundary
component contributes $\om$ to the background fermion number once, so
that we obtain 
\[
  \sum_{\al=1}^{\#_{insert}} F_\al = b~ \om \ .
\]
Taking into account that the current $G_{zz}$ is odd we find 
$\sum_{i=1}^n F_i + \sum_{a=1}^m F_a = m + (\om + 1) b$ for the
topological string amplitudes, where $F_i$ and $F_a$ are the fermion
numbers for bulk resp. boundary observables. 

Most of the subsequent formulas are conveniently expressed through the
introduction of a suspended grading, which we define by:
\beq
  \ll{suspgrade}
  \begin{array}{c@{~:=~}ll}
  \td i & F_i    & \quad\textrm{for a bulk field}~ \f_i \ ,\\
  \td a & F_a + 1& \quad\textrm{for a boundary field}~ \y_a \ ,
  \end{array}
\eeq
and $\td \om := \om + 1$. In other words, the suspended grading is the
fermion number of the topological descendent rather than the
topological observable itself. From now on we will refer to the
notions even and odd with respect to the suspended grading. The
$\bbZ_2$-selection rule takes the simple form 
\beq
  \ll{Z2select}
  \sum_{i=1}^n \td i + \sum_{l=1}^b \sum_{a_l=1}^{m_l} \td a_l = b~ \td \om \ .
\eeq

For most interesting models there is a close relation between the 
$\bbZ_2$-grading and the $U(1)$ charge. For Calabi--Yau compactifications this
is simple, since the $U(1)$ charge is the form degree and the 
fermion number is the form degree mod $2$. In Gepner models, on the
other hand, a relation is ensured in view of the orbifold action
\cite{Walcher04StabLG}. In particular, we have
$(\hc-3)$ mod $2 = \td\om$.

{\bf Remark:} For the rest of the paper we consider the case $\td\om =
0$, so that the $\bbZ_2$-selection rule is the same for arbitrary
numbers of boundaries $b$, \ie,
\[
  \sum_{i=1}^n \td i + \sum_{l=1}^b \sum_{a_l=1}^{m_l} \td a_l = 0 \ .
\]
In many practical situations we can concentrate on even bulk
fields. For instance, in Landau--Ginzburg models (not orbifolded!) the
bulk chiral ring includes only even fields. In the topological $A$-
and $B$-model the interest lies mainly on the marginal bulk operators,%
\footnote{We adopt the terminology of the untwisted $\cN = 2$ superconformal
  theory and call topological bulk and boundary observables with $q =
  q_L+q_R = 2$ resp. $q= 1$ \emph{marginal}. 
}
which are always even. Therefore, we subsequently consider only even
bulk fields.

\subsection{Symmetries, cyclic invariance, and unitality}

%% As last preparation for our computations in the next section let us
%% consider the symmetry properties of the topological string amplitudes.

In view of the topological nature of the amplitudes (\ref{allgb}), we
can deform the Riemann surface and exchange two 
boundary components. This leads to the graded symmetry:
\beq
  \ll{boundsymm}
  \cF^{b}_{A_1|\ldots|A_i|A_{i+1}|\ldots|A_b} =
  (-)^{\td A_i \td A_{i+1}}
  \cF^{b}_{A_1|\ldots|A_{i+1}|A_i|\ldots|A_b}
  \tfor \forall~ i=1,\ldots, b-1 \ ,
\eeq
where $\td A_i = \td a[i]_1+\ldots+\td a[i]_{m_i}$.

Moreover, the invariance of disk amplitudes under cyclic exchange of
boundary observables naturally extends to $b\geq 1$, \ie,
\beq
  \ll{cyclic}
  \cF^{b}_{A_1|\ldots|
    a[i]_1 a[i]_2 \ldots a[i]_{m_i}
    |\ldots|A_b} =
  (-)^{\td a[i]_1 (\td a[i]_2 + \ldots + \td a[i]_{m_i})} 
  \cF^{b}_{A_1|\ldots|
    a[i]_2 \ldots a[i]_{m_i} a[i]_1 
    |\ldots|A_b}\ .
\eeq

The behavior of the topological amplitudes under insertion of the
boundary unit operator $\unit$ was investigated in \cite{HLL04Ainfty}
and it was shown that all tree-level 
amplitudes $\cF^{0,1}_{a_1 \ldots a_m}$ with at least one unit
operator insertion vanish except for $m=3$. 
For all other amplitudes with $2g+b-2 \geq 0$ it is easy to see that
they vanish upon insertion of the unit, because the descendent of the
unit vanishes, \ie, $\unit^{(1)}=0$.%
\footnote{The only amplitude that does not vanish by this reasoning is
  the annulus with just the unit inserted on one boundary. It is,
  however, zero by the charge selection rule (\ref{chargeselect}).
} 
We define \emph{unitality} for the all-genus amplitudes
  $\cF^{b}_{A_1|\ldots |A_b}$ as the properties:%
\footnote{Strictly speaking, this is true if we deform the amplitudes
  by marginal bulk observables only. If we admit the full bulk chiral
  ring in the deformations there can also be non-vanishing
  bulk-boundary 2-point disk correlators with unit, \ie, the charge
  selection rule (\ref{chargeselect}) admits
  $\dl_i \cF^{0,1}_\unit = \bra \f_i~\unit\ket_{\Si_{0,1}} \neq 0$.
} 

\beq
  \ll{unital}
  \begin{array}{ll}
    \cF^{1}_{\unit a b} = (-)^{\td a} \rho_{ab}\\[10pt]
    \cF^{b}_{a[1]_1\ldots \unit \ldots a[1]_{m_1}|\ldots|A_b} = 0 & 
    \textrm{otherwise} \ ,
  \end{array}
\eeq
where 
\beq
  \ll{metric}
  \rho_{ab} = \bra ~\unit~ \y_a \y_b \ket_{\Si_{0,1}}
\eeq
is the topological open string metric. Note that the
charge selection rule (\ref{chargeselect}) for $\rho_{ab}$ reads
$\td a + \td b = 1$ (recall $\td\om =0$), and the metric is (graded)
symmetric 
\[
  \rho_{ab} = (-)^{\td a \td b} \rho_{ba} \ .
\]
Subsequently, we will use the topological open string metric
$\rho_{ab}$ rather than the symplectic structure 
$\om_{ab}=(-)^{\td a} \rho_{ab}$, which is commonly used in the
literature on cyclic $\cA_\infty$-structures \cite{GZ97OSFT,Kaj01SFT}.

\subsection{Special background charge $\hc=3$}
\ll{specialc3}

Topological string theories, for which the background charge $\hc$ is
equal to the critical dimension of the internal space of a
superstring compactification, \ie, $\hc=3$, have in many respects special
properties. The topological closed string at tree-level
is then governed by special geometry and the $tt^*$ equations; and
through the 'decoupling' of marginal operators from the relevant and
irrelevant ones the holomorphic anomaly equations of \cite{BCOV93KS}
take a particularly simple form.

In our situation $\hc=3$ is special in that the charge selection rule
(\ref{chargeselect}) is equal for arbitrary genus $g$ and number
of boundaries $b$, which leads to similar conclusions as for the topological
closed string \cite{BCOV93KS}. First of all, an all-genus amplitude
$\cF^{b}_{a_1\ldots a_m}(g_s;t)$, in which the fields satisfy the 
selection rule (\ref{chargeselect}), gets contributions from all
genera. This is not the case when $\hc\neq 3$, because then
$\cF^{g,b}_{a_1\ldots a_m}$ is non-vanishing for at most one
particular genus $\hat g$, \ie, 
\beq
  \nn
  \begin{array}{r@{~=~}ll}
    \cF^{b}_{A_1|\ldots |A_b} & 
    \sum_{g=0}^\infty g_s^{2g+b-2} \cF^{g,b}_{A_1|\ldots |A_b} & 
    \tfor \hc = 3,\\[5pt]
    \cF^{b}_{A_1|\ldots |A_b} & 
    g_s^{2\hg+b-2} \cF^{\hg,b}_{A_1|\ldots |A_b} & 
    \tfor \hc \neq 3, ~\textrm{and appr. genus}~\hg ,
  \end{array}
\eeq
where $\hg$ is determined by $\td A_1, \ldots, \td A_b$ and $b$
through the charge selection rule (\ref{chargeselect}). 

In fact, in many theories with $\hc\neq 3$ there is no solution to the
charge selection rule for $\hg \geq 1$ at all. For example, in a
topological string theory with $\hc=1$, such as the topologically
twisted non-linear sigma model on the torus or the associated
Landau--Ginzburg orbifold, the boundary observables have charges in the range 
$0 \leq q \leq 1$. Therefore, only Riemann surfaces with $\g_{g,b}
\geq 0$ give non-vanishing amplitudes. In particular, the annulus
and the torus amplitudes admit only marginal operator insertions.

\section{Quantum $\cA_\infty$-category}
\ll{sec:qAinfty}

As was shown in \cite{HLL04Ainfty} the topological open string
amplitudes satisfy a unital, cyclic $\cA_\infty$-algebra at 
tree-level:
\beq
  \ll{Ainfty}
  \sum_{l\leq k = 1}^{m} 
  (-)^{s_l} ~
  \cF^{0,1}_{a_1\ldots a_ l c a _{k+1}\ldots a_m}
  \rho^{cd} \cF^{0,1}_{d a_{l+1} \ldots a_k} = 0 \tfor m \geq 0 \ ,
\eeq
where $s_l = \td a_1 + \ldots + \td a_l$. Let us briefly recall how
this relation comes about. When we insert the BRST operator $Q=\oint_\cC
Q_z+\bQ_\bz$ in the disk amplitudes (\ref{disk3}) or
(\ref{disk11}) with the contour $\cC$ chosen such that it encloses
non of the operators then the amplitudes vanish. On the other hand,
if we deform the contour and act on all the bulk and boundary fields,
a series of contact terms gives rise to disks that bubble off
through a topological operator product and we eventually get the
$\cA_\infty$-structure (\ref{Ainfty}).
In this section we apply this idea to topological string
amplitudes of arbitrary genus $g$ and $b$ boundary components and show the
following result:
\bth
  \ll{thm:qAinfty}
  The all-genus topological string amplitudes 
  \[ \cF^{b}_{A_1|\ldots |A_b}(g_s;t) := 
    \sum_{g=0}^\infty g_s^{2g+b-2}~ 
    \cF^{g,b}_{A_1|\ldots |A_b}(t) \ ,
  \] 
  where $A_i$ is the collective index $a[i]_1 \ldots a[i]_{m_i}$,
  satisfy (what we call) the unital, cyclic quantum $\cA_\infty$-relations: 
  \begin{eqnarray}
    \ll{qAinftysum}
    \sum_{b'=1}^{b} \sum_{\s\in S_b} 
    \sum_{k\leq l = 1}^{m_{\s(1)}} \hspace*{-4mm}&&\hspace*{-4mm}
    \frac{(-)^{s_{\s(\td A)}+s_1}}{(b\!-\!b')!(b'\!-\!1)!} 
    \cF^{b-b'+1}_{a_1\ldots a_k c 
         a_{l\!+\!1}\ldots a_{m_{\s(1)}}|
         A_{\s(2)}|\ldots |A_{\s(b\!-\!b'\!+\!1)}}
    \rho^{cd} \cF^{b'}_{d a_{k\!+\!1} \ldots a_l|
                  A_{\s(b\!-\!b'\!+\!2)}|\ldots |A_{\s(b)}} \nn \\
    &&\hspace*{-25mm} =~
    \sum_{b'=1}^{b}  
    \sum_{k\leq l = 1}^{m_{b'}}
    (-)^{s_{b'}+ s_2} ~
    \rho^{cd} \cF^{b+1}_{a'_1\ldots a'_ k c a' _{l+1}\ldots a'_{m_{b'}}|
                           d a'_{k+1} \ldots a'_l|
			   A_1|\ldots \hat A_{b'} \ldots |A_b} \\
    &&\hspace*{-25mm} +~
    \sum_{b''<b'=1}^{b}  
    \sum_{k = 1}^{m_{b'}}\sum_{l = 1}^{m_{b''}}
    (-)^{s_{b'}+s_{b''}+ s_3} ~
    \rho^{cd} \cF^{b-1}_{a'_1\ldots a'_k c a''_{l+1} \ldots a''_{m_{b''}}
                   a''_1 \ldots a''_l d a' _{k+1}\ldots a'_{m_{b'}}|
		   A_1|\ldots \hat A_{b'}\ldots \hat A_{b''} \ldots |A_b}
    \nn \ ,
  \end{eqnarray}
  for $b \geq 1$, provided that $m_i \geq 1$ for $i=1,\ldots,b$. When
  $b=1$ the last term is zero. We call the quantum 
  $\cA_\infty$-relations \emph{weak} if both $\cF^{0,1}_{a}(t)$ and
  $\cF^{0,1}_{ab}(t)$ are non-vanishing, \emph{strong} if $\cF^{0,1}_{a}(t)=0$,
  and \emph{minimal} if $\cF^{0,1}_{a}(t)=\cF^{0,1}_{ab}(t)=0$. In the
  undeformed case, $t=0$, the quantum $\cA_\infty$-relations are minimal.
\eth
\begin{figure}[t]
  \begin{center}
    \epsfysize=2.8cm\centerline{\epsffile{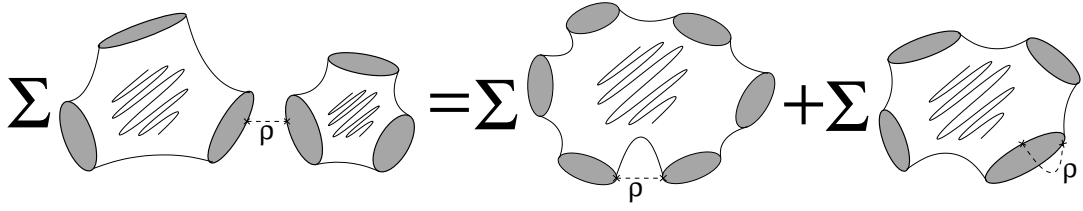}}
    \parbox{12truecm}{
      \caption{A diagrammatic representation of the quantum
      $\cA_\infty$-relations (\ref{qAinftysum}) for $b=5$. The
      expansion in $g_s$, \ie, the sum over all genera $g$, is
      indicated by the wiggly lines on the Riemann surfaces.
	\label{fig:qAinfty}} 
    }
  \end{center}
\end{figure}
The signs in (\ref{qAinftysum}) are:
\begin{eqnarray}
  \ll{signs}
  s_1 &=& \td a_1\!+\!\ldots\!+\! \td a_k \!+\! 
          (\td a_{l\!+\!1}\!+\!\ldots\!+\! \td a_{m_{\s(1)}} \!+\! 
           \td A_{\s(2)} \!+\! \ldots \td A_{\s(b\!-\!b'+\!1)})
	  (\td d\!+\!\td a_{k\!+\!1}\!+\!\ldots\!+\!\td a_l)
	  \nn \\
  s_2 &=& \td a'_1\!+\!\ldots\!+\! \td a'_k \!+\! 
          (\td a'_{l\!+\!1}\!+\!\ldots\!+\! \td a'_{m_{b'}})
          (\td d\!+\!\td a'_{k\!+\!1}\!+\!\ldots\!+\!\td a'_l)
          \\
  s_3 &=& \td a'_1\!+\!\ldots\!+\! \td a'_k \!+\! 
          (\td a''_1\!+\!\ldots\!+\! \td a''_l)
          (\td a''_{l\!+\!1}\!+\!\ldots\!+\!\td a''_{m_{b''}}) \!+\!
          \td A_{b''}
	  (\td d\!+\!\td a'_{k\!+\!1}\!+\!\ldots\!+\!\td a'_{m_{b''}})
          \nn \\
  s_{b'} &=& \td A_{b'}(\td A_1\!+\!\ldots\!+\!\td A_{b'-1}) \nn \ .
\end{eqnarray}
and $s_{\s(\td A)}$ is the Koszul sign for the permutation 
$\s \in S_n$ of boundary components with $\bbZ_2$-grading 
$\td A_i$. We used the abbreviations $a_j = a[\s(1)]_j$,
$a'_j=a[b']_j$ and $a''_j=a[b'']_j$. Fig. \ref{fig:qAinfty} shows a
pictorial representation of the quantum $\cA_\infty$-relations
(\ref{qAinftysum}). 

Cyclic invariance and unitality have been discussed earlier, so
that it remains to show formula (\ref{qAinftysum}). We do this in
several steps, starting with the insertion of the BRST operator in an
arbitrary Riemann surface $\Si_{g,b}$. This causes a factorization of
$\Si_{g,b}$ through degeneration channels where open or closed topological
observables are exchanged through an infinitely long throat. These
degenerations are described in terms 
of the boundary of the moduli space $\bar \cM_{g,b;n,m_1,\ldots,m_b}$ of a
bordered Riemann surface $\Si_{g,b}$ with $n$ (dressed) punctures in
the bulk and $m_i$ (dressed) punctures on the $i^{th}$ boundary. 
For a description of bordered Riemann surfaces and their moduli spaces
in terms of symmetric Riemann surfaces the reader may consult 
\cite{Sep91moduli}; see also \cite{KL01Enum,Liu02GWinv} for the context of
open Gromov--Witten invariants.

An important observation will be that all the
closed string degeneration channels vanish, provided that $m_i \geq 0$
for all $i=1,\ldots,b$. We will conclude that only the open string
factorization channels give rise to the quantum $\cA_\infty$-relations
(\ref{qAinftysum}).

\subsection{The boundary of the moduli space $\bar \cM_{g,b;0,m_1,\ldots,m_b}$}
\ll{sec:modulispace}

The proof of theorem \ref{thm:qAinfty} is more tractable if we start
with the undeformed boundary theory, which means that we do not insert
any bulk descendents in our amplitudes. After obtaining
(\ref{qAinftysum}) in this situation we will include bulk
deformations. 
%% Also, Genus $0$ amplitudes were
%% already considered in \cite{HLL04Ainfty}, therefore, we can
%% concentrate on higher genera. 

\begin{figure}[t]
  \begin{center}
    \epsfysize=4cm\centerline{\epsffile{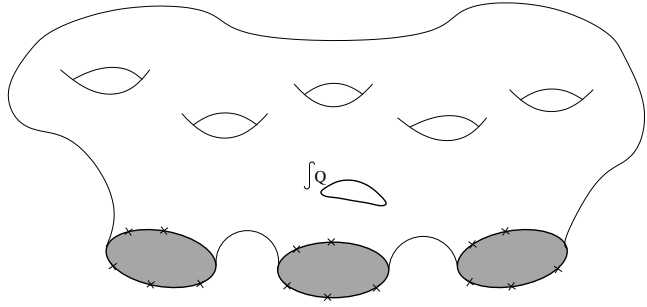}}
    \parbox{12truecm}{
      \caption{A genus 5 amplitude with three connected boundary
        component and a contour integral over the BRST current; the
        marked points indicate integrated boundary fields
        $\int\y_{a_i}^{(1)}$. The Beltrami differential with the
        fermionic currents (\ref{Beltrami}) are note depicted. 
	\label{fig:Qinsert}} 
    }
  \end{center}
\end{figure}

Our starting point is the amplitude
\beq
  \ll{Qinserted}
  \int_{\cM_{g,b}}  d^{6g+3b-6}m~
    \bra \big[Q, \int \Y_{A_1}\big|\ldots\big|\int \Y_{A_b}\big|
    \prod_{i=1}^{6g+3b-6} G_{m_i}\big] \ket_{\Si_{g,b}} = 0 
    \ , 
\eeq
for $2g+b-2 \geq 1$, $b \geq 1$, as depicted in Fig. \ref{fig:Qinsert}. Here, we used the
abbreviation
\beq
  \ll{intdomain}
  \int \Y_{A_i} = \int_{\cM^{i}_{m_i}} \Y_{A_i} = 
    P \int \y_{a[i]_1}^{(1)}(\ta_1) \ldots 
    \int \y_{a[i]_{m_i}}^{(1)}(\ta_{m_i}) \ .
\eeq
At this point some explanations on the moduli space in (\ref{Qinserted}) 
are in order. $\cM_{g,b}$ is the complex structure moduli space for
$\Si_{g,b}$. On an internal point of this moduli space, that is, for a
non-degenerate Riemann surface the boundary observables are integrated
in cyclic order, \ie, the integration domain in (\ref{intdomain}) is
\[\cM^{i}_{m_i} = (\De_{m_i-1}\rarrow S^1) \ .
\]
Here the simplex $\De_{m_i-1}$ is fibered over $S^1$ as follows: 
\[ \De_{m_i-1} = \left\{ (\tau_2,\ldots,\tau_{m_i}) \big| 
   \ta_1 < \ta_2 < \ta_3 < \ldots < \ta_{m_i} < \ta_1 + 2 \p \right\} \ ,
\]
where $\ta_1 \in S^1$ with $\ta_1 \simeq \ta_1 + 2 \p$. The total
moduli space for a Riemann surface with punctures on the boundary is
then a singular fibration  
\[\cM_{g,b;0,m_1,\ldots,m_b} =
(\cM^1_{m_1}\times\ldots\times\cM^b_{m_b}) \rarrow \cM_{g,b} \ .
\]

Having set up the basics about the moduli space, we investigate the
effect of the BRST operator  in (\ref{Qinserted}). Relations
(\ref{descendents}) and (\ref{Gdesc}) show that $Q$ gives rise to
a total derivative on the moduli space. Using
Stokes theorem the left-hand side of (\ref{Qinserted}) looks like
\[\int_{\dl\cM_{g,b;0,m_1,\ldots,m_b}} \bra \ldots \ket \ .
\]
The boundary of the moduli space
$\dl\cM_{g,b;0,m_1,\ldots,m_b}$ consists of singular configurations of
the Riemann surface. This requires a compactification of the moduli
space, $\overline\cM_{g,b;0,m_1,\ldots,m_b}$, meaning that
well-defined configurations of Riemann surfaces are added at the
singular locus of $\cM_{g,b;0,m_1,\ldots,m_b}$.
%%  so that the boundary is
%% well-defined. 

What kind of degenerations can occur when we deform the complex
structure of the Riemann surface $\Si_{g,b}$, \ie, what are the
boundary components of the moduli space, 
$\dl\overline\cM_{g,b;0,m_1,\ldots,m_b}$?
%% the $n$-dimensional boundary components $\De^{(n)}_m \subset \De_m$ for
%% $n=\{0,\ldots,m-1\}$, \ie, the vertices, edges, faces, ..., correspond
%% to collisions of $m+1-n$ fields.
To answer this let us divide the boundary configurations in three
major classes:\\[5pt] 
\emph{(i)} A sub-cylinder of $\Si_{g,b}$ becomes infinitely long and
thin, so that we can insert a complete system of topological closed
string observables.\\[5pt] 
\emph{(ii)} A sub-strip of $\Si_{g,b}$ constricts to be infinitely
long and we can insert a complete system of topological open string
fields.\\[5pt] 
\emph{(iii)} Several boundary fields come close together and we can
take the topological operator product, which once again amounts to
inserting a complete system of open string fields and bubbling of a
disk. (This degeneration could be included in \emph{(ii)}, but
we consider it separately for technical reasons, which become clear
below.)

In the subsequent sections we investigate these three situations
one-by-one and compute the resulting contributions to
(\ref{qAinftysum}).

\subsection{The closed string factorization channel}

\begin{figure}[t]
  \begin{center}
    \epsfysize=3.5cm\centerline{\epsffile{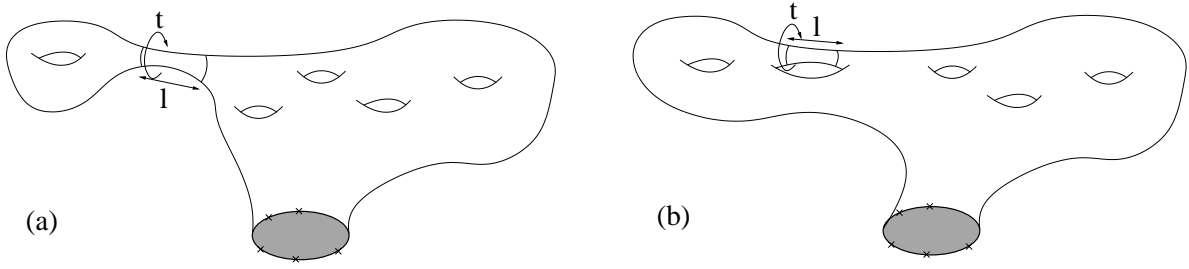}}
    \parbox{12truecm}{
      \caption{Closed string factorizations
	\label{fig:closedfact}} 
    }
  \end{center}
\end{figure}

A factorization in the closed string channel
(cf. Fig. \ref{fig:closedfact}) occurs if the BRST operator acts 
on one of the $G_{m_i}$'s and gives rise to the derivative $\der{m_i}$. 
For definiteness let us first consider the situation like in
Fig. \ref{fig:closedfact}a, where non of the factorization products is 
a disk. 

In the neighborhood of the degeneration, the Riemann
surface can be described in terms of the plumbing-fixture
procedure \cite{FS86AnalCFT,Pol88FactBos}. Take any two Riemann
surfaces, $\Si_1$ and $\Si_2$, and cut out a disk of radius
$\sqrt{|q|}$ on both of them, where $q \in \bbC$ and $|q|$ small. The
centers of the 
disks are located at $\hat z_1$ and $\hat z_2$. Let us parameterize
the neighborhood of the disks by the complex coordinates $z_1$ and
$z_2$. The plumbing-fixture procedure tells us 
to glue the Riemann surfaces through the transition function 
$z_2 - \hat z_2 = q/(z_1 - \hat z_1)$. This gluing describes a
cylinder of length $l$ and twist parameter $t$, determined by $q =
e^{i \ta} = e^{- l+i t}$. In the limit of infinite length, 
$l \rarrow \infty$, the tube connects at the points, $\hat z_1$ and
$\hat z_2$, to the Riemann surfaces, $\Si_1$ and $\Si_2$, respectively. 
Near this degenerate point the moduli space
$\bar \cM_{g,b;0,m_1\ldots m_b}$ can be parameterized by the coordinates 
$(\ta,\hat z_1, \hat z_2,\td m_i,\hat m_j)$. The moduli
$\td m_i$ and $\hat m_j$ are the moduli on the Riemann surfaces
$\Si_1$ and $\Si_2$. 

Instead of putting the moduli dependence on $(\ta,\hat z_1, \hat z_2)$
into the world sheet metric 
and using the Beltrami differentials $\m_i^z{}_\bz$, let us make a
local conformal transformation to a conformally flat metric on the
cylinder and describe the moduli dependence through the transition
function between the coordinates $z_1$ and $z_2$. An infinitesimal
change of the moduli can then be written in terms of the conformal
vector field 
\[  v^z(z_1-\hat z_1) = \de \hat z_1 v_1 + \de \ta v_\ta + \de \hat z_2 v_2
                      = \de \hat z_1 + \de \ta(z_1 -\hat z_1)
                      + \frac{\de \hat z_2}{e^{i\ta}}(z_1 -\hat z_1)^2 \ .
\]
The integrals over the Beltrami differentials become
\beq 
  \ll{tubeBeltrami}
  \int G_{m_i} = \oint_\cC G_{zz} v^z_i + \bG_{\bz \bz} \bar v^\bz_i
\eeq
for $i = 1,2,\ta$. The cycle $\cC$ wraps once around the tube.

The degeneration that we described here corresponds to the situation
when $Q$ acts on $G_l$ and gives rise to the total derivative
$\der{l}$. Using (\ref{tubeBeltrami}) the amplitude on the
degenerate Riemann surface becomes
\bea
   \ll{closedfact}
   &&\lim_{l\rarrow \infty} \int\bra \ldots 
   [G_{-1},[\bG_{-1},\f_m(\hat z_1)]] 
   \hspace*{-3mm}\prod_{i=1}^{6g'+3b'-6} \hspace*{-3mm} G_{\td m_i}
   \ket_{\Si_{g',b'}}~ 
   \eta^{mn}~ \times \\ \nn
   &\times & \braarg{\f_n}  
   q^{L_0} \bar q^{\bar L_0} \ketarg{\f_k} \eta^{kl}~
   \int\bra [i(G_{0}-\bG_{0}), [G_{-1}, [\bG_{-1}, \f_l(\hat z_2)]]]
   \ldots 
   \hspace*{-4mm}\prod_{j=1}^{6g''+3b''-6}\hspace*{-4mm} G_{\hat m_j} 
   \ket_{\Si_{g'',b''}} \ ,
\eea
where $g'+g''=g$, $b'+b''=b$. We omitted the details about the
integration over the moduli spaces and also about the boundary observables,
which are indicated by dots. In the limit $l\rarrow \infty$, $q$
vanishes and only states with $L_0=\bar L_0 = 0$, \ie, the topological
closed string observables survive, so that $\eta^{mn}$ restricts to the
inverse of the topological closed string metric.

The important point is then that the zero mode $G_0\!-\!\bG_{0}$ remains in
(\ref{closedfact}) and acts on the bulk observable, so that the whole
expression vanishes by the gauge condition (\ref{gauge}).
A similar argument applies to the factorization channel of
Fig. \ref{fig:closedfact}b, which therefore vanishes too. 

Recall that we excluded so far the situation where one of the two
Riemann surfaces, $\Si_1$ or $\Si_2$, is a disk. Let us consider this
case now. Other than before there is no twist parameter $t$, so that the
factorization becomes
\bea
   \ll{closedfact2}
   \int\bra \ldots 
   [G_{-1}, [\bG_{-1}, \f_m(\hat z_1)]] 
   \hspace*{-3mm}\prod_{i=1}^{6g+3b-9} \hspace*{-3mm} G_{\td m_i}
   \ket_{\Si_{g,b-1}}~ 
   \eta^{mn}~
   \bra  \f_n(\hat z_2) \int \Psi_{A_i}
   \ket_{\Si_{0,1}}
   \ .
\eea
Notice that $i(G_{0}-\bG_{0})$ does not appear because of the absence
of the twist parameter $t$,
and there is no $[G_{-1},[\bG_{-1},\f_l]]$ associated to the modulus $\hat
z_2$ either, which reflects the fact that the disk has a conformal
Killing vector field that can be used to fix $\hat z_2$.
Let us use now our assumption that we have at least one insertion of
an observable on each boundary, \ie, $m_i \geq 1$. In
\cite{HLL04Ainfty} it was then shown that a disk amplitude like the one in
(\ref{closedfact2}) vanishes in view of a conformal Ward identity.%
\footnote{A similar Ward identity is responsible for the fact that the
  topological metric $\rh_{ab}$ does not get deformations.
}
We conclude that, for $m_i \geq 1$,
\emph{factorizations in the closed string channel do not contribute at
all} to the quantum $\cA_\infty$-structure (\ref{qAinftysum}).

\subsection{Boundaries without observables -- non-stable configurations}

Let us briefly comment on the case $m_i = 0$. The expression
(\ref{closedfact2}) becomes 
\beq
  \ll{non-stable}
   \int\bra \ldots 
   [G_{-1},[\bG_{-1},\f_m(\hat z_1)]] 
   \hspace*{-3mm}\prod_{i=1}^{6g+3b-9} \hspace*{-3mm} G_{\td m_i}
   \ket_{\Si_{g,b-1}}~ 
   \eta^{mn}~
   \bra  \f_n(\hat z_2) \ket_{\Si_{0,1}}
   \ ,
\eeq
which corresponds to a non-stable configuration, because the conformal
Killing vector that rotates the disk is not fixed.
The simplest example for such a situation is the factorization of the
annulus amplitude,% 
\footnote{The charge selection rule (\ref{chargeselect}) tells us that
  the single observable on the boundary must be the identity operator
  $\unit$.
}
\ie,
\beq
  \ll{annoinsert}
  \int_0^\infty \!\!dL ~
  \bra \left[ Q,~ \unit~ | ~.~ |~ G_L \right] \ket_{\Si_{0,2}}  = 0 \ .
\eeq
The open string factorization channel
gives the Witten index, or intersection number,
\beq
  \ll{Windex}
  \sum_{a,b} \cF_{\unit ab}\rho^{ab} = -\Tr_{\cH^{\al\be}_{op}} (-)^F \ ,
\eeq
where we used the equation in the unitality properties (\ref{unital}) that
relates 3-point functions to the topological metric. The closed string
channel gives 
\[  \bra \f_m~ \unit \ket_{\Si_{0,1}}~ \et^{mn} \bra \f_n \ket_{\Si_{0,1}} \ ,
\]
so that the factorization of the annulus amplitude (\ref{annoinsert})
can be interpreted as topological Cardy relation of 2d topological
field theory \cite{MooreSegal,Laz00TFT}. In general, one should be
cautious about considering factorizations that involve
non-stable configurations like (\ref{non-stable}). They
indicate ambiguities related to divergences in topological
amplitudes (cf. \cite{HLN06torus}).

\subsection{The open string factorization channels}

Let us turn now to the non-vanishing contributions to the quantum $\cA_\infty$
relations (\ref{qAinftysum}), which come from open string
factorization channels. Factorizations in the open string channel
corresponding to an infinitely long strip are shown in
Fig. \ref{fig:openfact}.

\subsubsection*{The left-hand side of the quantum
      $\cA_\infty$-relation (\ref{qAinftysum})}

\begin{figure}[t]
  \begin{center}
    \epsfysize=4cm\centerline{\epsffile{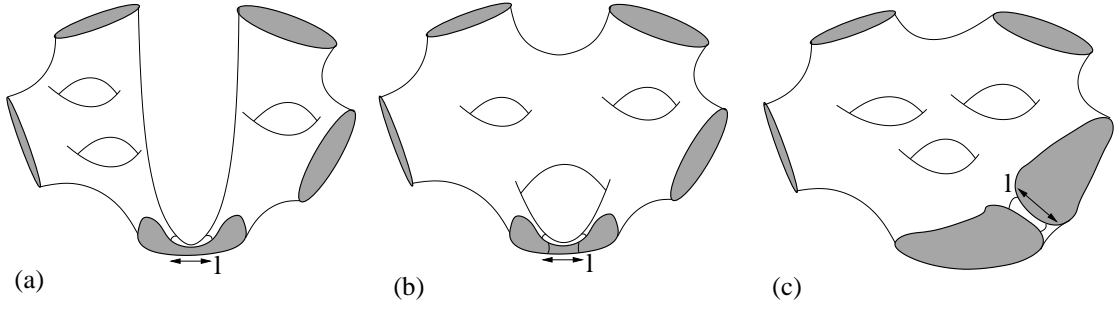}}
    \parbox{12truecm}{
      \caption{Open string factorizations through  infinitely long
      strips. \label{fig:openfact}} 
    }
  \end{center}
\end{figure}

We consider the situation in Fig. \ref{fig:openfact}a first. Locally
near the degeneration point the moduli space can be parameterized by
%the coordinates 
$(l,\hat x_1,\hat x_2,\td m_1,\ldots,\td m_{3|\td \g|},
\hat m_1,\ldots,\hat m_{3|\hat \g|})$. The first
three coordinates parameterize the length $l$ of the strip as well as
the positions $\hat x_1$ and $\hat x_2$ of the punctures, where the
strip ends on the surface boundaries. $\td m_i$ and $\hat m_j$ are the
moduli of the resulting Riemann surfaces $\Si_{g',b'}$ and $\Si_{g'',b''}$,
respectively. Here, $g'+g''=g$ and $b'+b''=b+1$. Since $Q$ acted on
$G_l$ we have to evaluate $l$ at infinity. The Beltrami differentials
associated to the moduli $\hat x_1$ and $\hat x_2$ localize around the
punctures as before. The channel in Fig. \ref{fig:openfact}a gives 
\beq
  \ll{openfactA}
    (-)^{3|\td \g|} \lim_{l\rarrow\infty} 
    \bra \int \Y_{A_1}\ldots\int \Y_{A_b}\big|
    \int_{\td \cM_{g',b'}}\hspace*{-8mm}  d^{3|\td \g|}\td m~
    \prod_{i=1}^{3|\td \g|} G_{\td m_i} 
    \int_{\hat x_1} G_{\hat x_1} 
    \int_{\hat x_2} G_{\hat x_2}
    \int_{\hat \cM_{g'',b''}}\hspace*{-8mm}  d^{3|\hat \g|}\hat m~
    \prod_{i=1}^{3|\hat \g|} G_{\hat m_i} 
\ket \ .
\eeq
The sign comes from pulling $Q$ through all the operators. We used 
$\td A_1 + \ldots + \td A_b = 1$. Notice that contributions with 
$(g',b')=(0,1)$ or $(g'',b'')=(0,1)$ vanish, because the degeneration
results into disk amplitudes that vanish by a conformal Ward identity
\cite{HLL04Ainfty}.  

We have not decided yet, which boundary component, that is to say,
which fields $\Y_{A_i}$ are involved in the factorization. Let us pick
$\Y_{A_1}$ first and take care of all other boundaries afterwards. 
The boundary observables are, after factorization, split into a
collection $\y^{(1)}_{a_{l+1}} \ldots \y^{(1)}_{a_{k}}$ on
$\Si_{g'',b''}$ and 
$\y^{(1)}_{a_{k+1}} \ldots \y^{(1)}_{a_{m_1}}
 \y^{(1)}_{a_{1}}\ldots \y^{(1)}_{a_{l}}$ on $\Si_{g',b'}$, where 
$l,k = 1,\ldots,m_1$ with $l \leq k$. In order to avoid over-counting
 we take $\y^{(1)}_{a_{1}}$ always to be on $\Si_{g',b'}$.

There is another choice that determines how the
remaining fields $\Y_{A_i}$ for $i=2,\ldots,b$ are distributed among 
$\Si_{g',b'}$ and $\Si_{g'',b''}$. We pick again the simplest choice: 
$\Si_{g',b'}$ carries the fields $\Y_{A_i}$ for $i=2,\ldots,b'$ and
$\Si_{g'',b''}$ carries $\Y_{A_i}$ for $i=b'+1,\ldots,b$.

Reshuffling the operators and inserting a complete system of boundary
observables gives
\bea
  \nn (-)^s \rho^{cd} \hspace*{-7mm}&&
  \bra \int \y^{(1)}_{a_{l+1}} \ldots \y^{(1)}_{a_{k}} \big|
       \int \Y_{A_2} \ldots \int \Y_{A_{b'}} \big|
       \int_{\td \cM_{g',b'}}\hspace*{-8mm}  d^{3|\td \g|}\td m~
       \prod_{i=1}^{3|\td \g|} G_{\td m_i} 
       \int_{\hat x_1} G_{\hat x_1} \y_c(\hat x_1) \ket_{\Si_{g',b'}} \times \\
  \nn \times&&\hspace*{-5mm} 
       \bra \y_d(\hat x_2) \int \y^{(1)}_{a_{k+1}} \ldots \y^{(1)}_{a_{l}} \big|
       \int \Y_{A_{b'+1}} \ldots \int \Y_{A_b} \big|
       \int_{\hat x_2} G_{\hat x_2}
       \int_{\td \cM_{g'',b''}}\hspace*{-8mm}  d^{3|\hat \g|}\hat m~
       \prod_{i=1}^{3|\hat \g|} G_{\hat m_i} 
  \ket_{\Si_{g'',b''}} \ .
\eea
The sign is 
$s=1\!+\!(3|\td \g|\!+\!1)\td d \!+\! (\td a_1\!+\!\ldots\!+\!\td a_l)
(\td a_{l\!+\!1}\!+\!\ldots\!+\!\td a_{m_1}) \!+\! 
(\td A_2 \!+\! \ldots \!+\! \td A_{b'})
(\td a_{k\!+\!1}\!+\!\ldots\!+\!\td a_l)$.

Now we use the fact that the Beltrami differentials associated to the
positions $\hat x_i$ localize around the punctures:  
$\int_{\hat x_1} G_{\hat x_1} \y_c(\hat x_1) \rarrow \int_{\hat x_1}
\y_c^{(1)}$. Further reshuffling of fields and using the definition of
topological amplitudes in (\ref{highergb}) we obtain
\beq
  \ll{openfact1}
  (-)^{s_1} \rho^{cd} 
  \cF^{g'',b''}_{a_1\ldots a_k c a_{l+1}\ldots a_{m_1}|A_2|\ldots|A_{b'}}
  \cF^{g',b'}_{d a_{k+1}\ldots a_l|A_{b'+1}|\ldots|A_b} \ ,
\eeq
where 
\[
  s_1 = \td a_1\!+\!\ldots\!+\! \td a_k \!+\! 
        (\td a_{l\!+\!1}\!+\!\ldots\!+\! \td a_{m_{1}} \!+\! 
        \td A_{2} \!+\! \ldots \td A_{b'})
	(\td d\!+\!\td a_{k\!+\!1}\!+\!\ldots\!+\!\td a_l) \ .
\]

In order to take into account  the factorizations involving the other boundary
components and all the inequivalent distributions of remaining fields
$\int \Y_{A_i}$, we exchange the observables $\int \Y_{A_i}$ according
to an element $\s$ in the symmetric group $S_b$. This gives rise to
the Koszul sign $(-)^{s_{\s(\td A)}}$. Then, summing up all factorization
channels gives
\bea
  \ll{sumopenfact1}
    \sum_{g'=0}^{g}{}^{'} \sum_{b'=1}^{b}{}^{'} 
    &&\hspace*{-10mm}
    \sum_{\s\in S_b} \sum_{l\leq k = 1}^{m_{\s(1)}} 
    \frac{(-)^{s_{\s(\td A)}+s_1}}{(b\!-\!b')!(b'\!-\!1)!} \times \\
    &\times&
    \cF^{g',b-b'+1}_{a_1\ldots a_l c 
         a_{k\!+\!1}\ldots a_{m_{\s(1)}}|
         A_{\s(2)}|\ldots |A_{\s(b\!-\!b'\!+\!1)}}
    \rho^{cd} \cF^{g-g',b'}_{d a_{l\!+\!1} \ldots a_k|
                  A_{\s(b\!-\!b'\!+\!2)}|\ldots |A_{\s(b)}} \nn \ .
\eea
The factor $[(b\!-\!b')!(b'\!-\!1)!]^{-1}$ accounts for
over-counting, and $\sum{}^{'}$ means that $(g',b')=(0,b)$ and
$(g',b')=(g,1)$ are not included in the sum. 
In fact, the latter contributions come from class \emph{(iii)} in
our list of degenerations in section \ref{sec:modulispace}. Let us briefly
consider those before we proceed to Fig. \ref{fig:openfact}b.

When $Q$ acts on the integrated descendents $\int\y_{a}^{(1)}$, it acts as a
boundary operator on one of the fiber components $\cM^{i}_{m_i}$ of
the moduli space $\bar \cM_{g,b;0,m_1\ldots m_b}$. The boundary of
$\cM^{i}_{m_i}$ corresponds to situations where two or more
observables collide. So we have to sum over all possible contact terms
of boundary observables. 
This is exactly the same effect that gave rise to the (classical)
$\cA_\infty$-structure for disk amplitudes in \cite{HLL04Ainfty}. If,
for instance, the fields $\y_{a_{l+1}}$ through $\y_{a_k}$ for
$l,k=1,\ldots,m_i$ and $l+2\leq k$, come
together very closely a disk with these fields bubbles off and we get
\bea
  \ll{openfactC1}
  -(-)^{\td a_1 + \ldots + \td a_{k-1}} &&\hspace{-5mm}
  \bra \y_{a_1}^{(1)}\ldots \y_{a_l}^{(1)} \y_c^{(1)}
  \y_{a_{k+1}}^{(1)}\ldots  \y_{a_{m_1}}^{(1)} | \int \Y_{A_2} \ldots| 
  \prod_{j=1}^{6g+3b-6} G_{\td m_j}\ket_{\Si_{g,b}} \times  \nn \\
  &&\times~ \om^{cd}
  \bra \y_d \y_{a_{l+1}} \y_{a_{l+2}}^{(1)}\ldots 
  \y_{a_{k-1}}^{(1)} \y_{a_k} \ket_{\Si_{0,1}}\ ,
\eea
and similarly
\bea
  \ll{openfactC2}
  -(-)^{s} &&\hspace{-5mm}
  \bra \y_{a_1} \y_{a_2} \y_{a_3}^{(1)} \ldots 
  \y_{a_l}^{(1)} \y_c^{(1)} \y_{a_{k+1}}^{(1)}\ldots 
  \y_{a_{m_1-1}}^{(1)} \y_{a_{m_1}} \ket_{\Si_{0,1}} \times \\
  &&\times~ \om^{cd}
  \bra  \y_d^{(1)} \y_{a_{l+1}}^{(1)}\ldots \y_{a_k}^{(1)}  
  | \int \Y_{A_2} \ldots| 
  \prod_{j=1}^{6g+3b-6} G_{\td m_j}\ket_{\Si_{g,b}} \nn \ ,
\eea
where $s = \td a_{m_1} +\td a_2 + \ldots + \td a_l + (\td a_{k+1} + \ldots + \td
a_{m_1})(\td d + \td a_{l+1} + \ldots + \td a_k)$.
When we compare these expressions with (\ref{openfact1}) taking into
account the sign in (\ref{disk3}), we see immediately that they
provide exactly the two missing terms with $(g',b')=(0,b)$ and
$(g',b')=(g,1)$ in (\ref{sumopenfact1}).

The all-genus amplitudes (\ref{allgb}) allow us to combine the
factorizations that we have studied so far into the
left-hand side of (\ref{qAinftysum}). 
%% Notice, however, that we are
%% still in the undeformed situation, \ie, the amplitudes are
%% $\cF^{b}_{A_1\ldots A_b}|_{t=0}$, in particular,
%% $\cF^{0,1}_a\big|_{t=0} = \cF^{0,1}_{ab}\big|_{t=0} = 0$.

\subsubsection*{The right-hand side of the quantum
      $\cA_\infty$-relation (\ref{qAinftysum})}

The channel shown in Fig. \ref{fig:openfact}b gives a
degeneration resulting in a single Riemann surface $\Si_{g-1,b+1}$, that
is, one boundary component splits into two, thus increasing the number
of boundaries by one and decreasing the genus by one. We rearrange
the boundary components in the 
amplitude such that the observables $\Y_{A_{b'}}$, which are effected
by the degeneration, are at the first position. This gives rise to the sign 
$s_{b'}$ defined in (\ref{signs}). The observables are
split into $\y^{(1)}_{a'_{l+1}}\ldots \y^{(1)}_{a'_k}$ and 
$\y^{(1)}_{a'_{k+1}}\ldots \y^{(1)}_{a'_{m_{b'}}}
 \y^{(1)}_{a'_{1}}\ldots \y^{(1)}_{a'_l}$. Here, $a'_i = a[b']_i$.
In the limit $l \rarrow \infty$ the amplitude becomes
\bea
  \nn
  (-)^{s+s_{b'}} \rho^{cd} 
  \bra \y_c(\hat x_1) \int \y^{(1)}_{a'_{k+1}} \ldots \y^{(1)}_{a'_l} 
  \hspace*{-5mm}&\big|&\hspace*{-5mm}
  \int \y^{(1)}_{a'_{l+1}}\ldots \y^{(1)}_{a'_k} \big|
  \int \Y_{A_1} \ldots
%%       \widehat{\Y_{A_{b'}}} \ldots \int \Y_{A_b} 
  \big| \y_d(\hat x_2) \times \\
  \nn 
  &\times& \int_{\hat x_1}\!\! G_{\hat x_1}\!\! 
  \int_{\hat x_2}\!\! G_{\hat x_2}\!\! 
  \int_{\td \cM_{g-1,b+1}} \hspace*{-8mm} d^{3|\td \g|}\td m
  \prod_{j=1}^{3 |\td \g|} G_{\td m_j} 
  \ket_{\Si_{g-1,b+1}} \ ,
\eea
where 
$s=(\td a'_{l+1}+\ldots + \td a'_{m_{b'}})(\td a'_1 + \ldots + \td a'_l)$.

We commute $\y_c$ and $\y_d$ through the other fields to their 'right'
positions and make use of the localization
$\int_{\hat x_1} G_{\hat x_1} \y_c \rarrow \int_{\hat x_1}
\y_c^{(1)}$. After further 
reshuffling of the observables and using (\ref{highergb}) we obtain: 
\[
  -(-)^{s_{b'}+s_2}
  \rho^{cd}~ 
  \cF^{g-1,b+1}_{a'_1\ldots a'_l c a'_{k+1}\ldots a'_{m_b'}|
    d a'_{l+1} \ldots a'_k | A_1| \ldots| \hat A_{b'}| \ldots| A_b}
%%   \bra  \y_{a_0}^{(1)}\ldots \y_{a_l}^{(1)} \y_{c}^{(1)}
%%   \y_{a_{k+1}}^{(1)} \ldots \y_{a_m}^{(1)}  \big|  
%%   \y_{d}^{(1)} \y_{a_{l+1}}^{(1)} \ldots \y_{a_k}^{(1)}  \big| 
%%   \int \Y_{A_1} \ldots \widehat{\Y_{A_{b'}}} \ldots \int \Y_{A_b} \big|
%%   \int_{\td \cM_{g,b+1}} \hspace*{-8mm} d^{3|\td \g|}\td m
%%   \prod_{j=1}^{3 |\td \g|} G_{\td m_j} 
%%   \ket 
  \ , 
\]
where the signs can be found in (\ref{signs}). Summing over all such
channels yields
\beq
  \ll{sumopenfact2}
  -\sum_{b'=1}^b \sum_{l\leq k = 1}^{m_{b'}} (-)^{s_{b'}+s_2} 
  \rho^{cd}~ 
  \cF^{g-1,b+1}_{a'_1\ldots a'_l c a'_{k+1}\ldots a'_{m_b'}|
    d a'_{l+1} \ldots a'_k | A_1| \ldots| \hat A_{b'}| \ldots| A_b}
  \ , 
\eeq
which provides the (undeformed) first term on the right-hand side of
(\ref{qAinftysum}).

Finally we have to look at the degeneration in
Fig. \ref{fig:openfact}c, where the genus $g$ stays the same and
two boundary components join into one. Let us pick the fields
$\Y_{A_{b''}}$ and $\Y_{A_{b'}}$ on the colliding boundaries, where
$b',b'' = 1,\ldots,b$ and $b''< b'$. Pulling these observables
through the other fields to the first two positions in the amplitude
gives the sign 
$s_{b''}+s_{b'}$. In the degeneration limit we insert the complete
system $\rho^{cd}\y_c \y_d$ in such a way that $\y_c$ is located
between $\y^{(1)}_{a'_k}$ and $\y^{(1)}_{a''_{l+1}}$, whereas $\y_d$
is located between $\y^{(1)}_{a''_l}$ and $\y^{(1)}_{a'_{k+1}}$. Here
$k = 1,\ldots,m_{b'}$ and $l = 1,\ldots,m_{b''}$. We obtain:
\bea
  \nn
  (-)^{s+s_{b'}+s_{b''}} \rho^{cd} 
  \bra \y_c(\hat x_1)   
  &&\hspace*{-10mm}
  \int \y^{(1)}_{a'_{k+1}} \ldots \y^{(1)}_{a'_{m_{b'}}}
       \y^{(1)}_{a'_{1}} \ldots \y^{(1)}_{a'_k} 
  \int \y^{(1)}_{a''_{l+1}} \ldots \y^{(1)}_{a'_{m_{b''}}}
       \y^{(1)}_{a''_{1}} \ldots \y^{(1)}_{a''_l}  
  \big|  \times \\
  \nn 
  &\times& \int \Y_{A_1} \ldots
%%       \widehat{\Y_{A_{b'}}} \ldots \int \Y_{A_b} 
  \big| \y_d(\hat x_2)
\int_{\hat x_1}\!\! G_{\hat x_1}\!\! 
  \int_{\hat x_2}\!\! G_{\hat x_2}\!\! 
  \int_{\td \cM_{g,b-1}} \hspace*{-8mm} d^{3|\td \g|}\td m
  \prod_{j=1}^{3 |\td \g|} G_{\td m_j} 
  \ket_{\Si_{g,b-1}} \ ,
\eea
where $\y_c$ and $\y_d$ are not yet at the positions according to
their boundary condition labels. The sign is
$s\!=\!(\td a'_1 \!+\!\ldots \!+\! \td a'_k)
   (\td a'_{k\!+\!1} \!\!+\!\ldots \!+\! \td a'_{m_{b'}}) \!+\! 
   (\td a''_1 \!+\!\ldots \!+\! \td a''_l)
   (\td a''_{l\!+\!1} \!\!+\!\ldots \!+\! \td a''_{m_{b''}})$.
Following the same steps as for the other factorizations and
collecting all contributions we obtain:
\beq
  \ll{sumopenfact3}
  -\sum_{b''<b'=1}^b \sum_{k = 1}^{m_{b'}} \sum_{l = 1}^{m_{b''}}
  (-)^{s_{b'}+s_{b''}+s_3} \rho^{cd}
  \cF^{g,b-1}_{a'_1\ldots a'_k c a''_{l+1} \ldots a''_{m_{b''}} a''_1
    \ldots a''_l d a'_{k+1} \ldots a'_{m_{b'}}|
    A_1\ldots \hat A_{b''}\ldots \hat A_{b'}\ldots A_b } \ .
\eeq
This provides the final contribution to the quantum $\cA_\infty$
relation (\ref{qAinftysum}). Actually, what we have found so far are
the undeformed, minimal quantum $\cA_\infty$-relations for the
undeformed amplitudes $\cF^b_{A_1|\ldots|A_b}(g_s;t=0)$, in particular, 
$\cF^{0,1}_{a}(t=0) = \cF^{0,1}_{ab}(t=0) = 0$. 

{\bf Remarks:} Observe that the restriction $m_i=0$ for $i=1,\ldots,b$
did not play any r\^ole in the analysis of the open string factorization
channels. This means that in situations with 'bare' boundary
components, the quantum $\cA_\infty$-relations (\ref{qAinftysum}) hold
only up to non-stable configurations like (\ref{closedfact2}). 

Strictly speaking, we are not  done with the factorizations of the
undeformed amplitudes yet, because the annulus amplitude, equation
(\ref{annulus1}) with $Q$ insertion and $t=0$, was not included in our
considerations so far. We just state here that the gymnastics of the
previous section can be applied  as well and leads to the still
missing terms in (\ref{qAinftysum}).

\subsection{Including closed string deformations}

The inclusion of closed string deformations, \ie, insertions of bulk
observables in the amplitudes, has a quite trivial effect. First of all,
contact terms between bulk fields do not contribute if the
regularization is chosen appropriately (cf. \cite{HLL04Ainfty}).%
\footnote{Another way to say that is that the minimal $L_\infty$
  structure for (closed) topological string is trivial, \ie, all
  $L_\infty$ brackets vanish; see \cite{KS05OCHA}.
}

Suppose we insert one (integrated) bulk descendent in the
amplitude (\ref{Qinserted}). If $Q$ acts on boundary fields or
the $G_{m_i}$'s we obtain the same factorization channels as
before.  In the case (\ref{openfact1}), where the Riemann surface
splits into two, $\Si_1$ and $\Si_2$, the integration over the bulk
observable splits too, \ie, $\int_{\Si} \f_i^{(2)} \rarrow \int_{\Si_1} 
\f_i^{(2)}+\int_{\Si_2}\f_i^{(2)}$. Notice that this is consisted with,
and therefore allows, the formal integration of bulk descendents to
the deformed amplitudes (\ref{highergb}). 

If $Q$ acts, on the other hand, on the bulk descendent we
get contact terms between this bulk descendent and boundary fields, which gives
rise to disks that bubble off the Riemann surface
(cf. \cite{HLL04Ainfty}). This provides additional contributions for
the quantum $\cA_\infty$-relations involving
$\cF^{0,1}_a(t)$ and $\cF^{0,1}_{ab}(t)$.

We conclude that \emph{the closed string observables deform the
minimal quantum $\cA_\infty$-structure for undeformed amplitudes into
the weak quantum $\cA_\infty$-structure (\ref{qAinftysum}) for
deformed amplitudes.}

\section{Comments on the quantum $\cA_\infty$-relations}
\ll{sec:comments}

So far we have neglected the boundary condition
labels, $\al,\ldots$ for convenience. Reintroducing them makes apparent that
relation (\ref{qAinftysum}) defines a cyclic, unital quantum $\cA_\infty$
category (rather than an algebra). It is the quantum
version of a (classical) $\cA_\infty$-category, which was originally
introduced in \cite{Fukaya93AinftyCat}. The boundary conditions (or D-branes)
are the objects, $\al,\be,\ldots \in Obj(\Acat)$, and the boundary observables
are the morphisms, $\y^{\al\be}_a \in Hom_\Acat(\al,\be) = \cH^{\al\be}_{op}$. 
The formulation of the classical $\cA_\infty$-relations in terms of scattering
products $r_n : \cH^{\al_1\al_2}_{op} \otimes \ldots 
\otimes \cH^{\al_n\al_{n\!+\!1}}_{op} \rarrow \cH^{\al_1\al_{n\!+\!1}}_{op}$
can be found in the literature. We refer to the recent review
\cite{KS05OCHA} on this subject, and references therein. 

Instead of elaborating on this issue we want to focus subsequently on the
r\^ole and the effects of the charge selection rule on the quantum
$\cA_\infty$-relations (\ref{qAinftysum}) in models with $\hc =3$.
For this purpose let us distinguish between boundary condition preserving
observables (BPO) $\y^{\al\al}_a \in \cH^{\al\al}_{op}$ and boundary condition
changing observables (BCO) $\y^{\al\be}_a \in \cH^{\al\be}_{op}$ for 
$\al \neq \be$.

Consider a single D-brane so that we have only BPOs. We assume to be in a
model where the boundary condition 
preserving sector has only integral $U(1)$ charges, \ie, $q=0,1,2,3$, and the
unique observable of charge $0$ is the unit operator $\unit$. This is
the case in most models of interest.

It was pointed out in \cite{AK04Superpot} that the disk amplitudes then have a
particularly simple form and that the (classical) $\cA_\infty$-relations are
trivially satisfied. A similar argument can be adopted to the all-genus
topological string amplitudes $\cF^{b}_{A_1 |\ldots| A_b}$ and goes as
follows: Recall first that for $\hc = 3$ the charge selection rule
(\ref{chargeselect}) is the same irrespective of the Euler character of the
Riemann surface. In particular, the selection rule (\ref{chargeselect}) admits
to insert \emph{only} 
marginal boundary observables in the amplitude $\cF^{b}_{A_1|\ldots|A_b}$ and
moreover an arbitrary number of them. Take such an amplitude and substitute
one of the marginal observables by a charge $2$ (or $3$) one. In order to
obtain a non-vanishing amplitude the selection rule
(\ref{chargeselect}) forces us to introduce one (or two) units
$\unit$ in the amplitude. On the other hand, by the unitality property
(\ref{unital}) the only non-vanishing amplitudes with unit are disk 3-point
correlators. We conclude that amplitudes on a single D-brane (with  
the above assumptions) \emph{(i)} have only marginal insertions or \emph{(ii)}
are given by $\cF^{0,1}_{\unit ab}$.
From this observation it follows readily that \emph{the quantum
$\cA_\infty$-relations (\ref{qAinftysum}) are trivially satisfied when
we consider a single D-brane.}

Only in multiple D-brane situations, that is for a quantum
$\cA_\infty$-category, the algebraic equations (\ref{qAinftysum}) give
non-trivial relations and can be used as constraints on the
amplitudes. In fact, they provide a means of determining higher genus
multiple-boundary amplitudes $\cF^{g,b}_{A_1|\ldots|A_b}$ recursively from 
amplitudes with larger Euler character. To see this let us rewrite
(\ref{qAinftysum}) in such a way that they look diagrammatically 
as follows: 
\beq
    \ll{recursive}
    \parbox{4.2cm}{\epsfxsize=4cm \epsffile{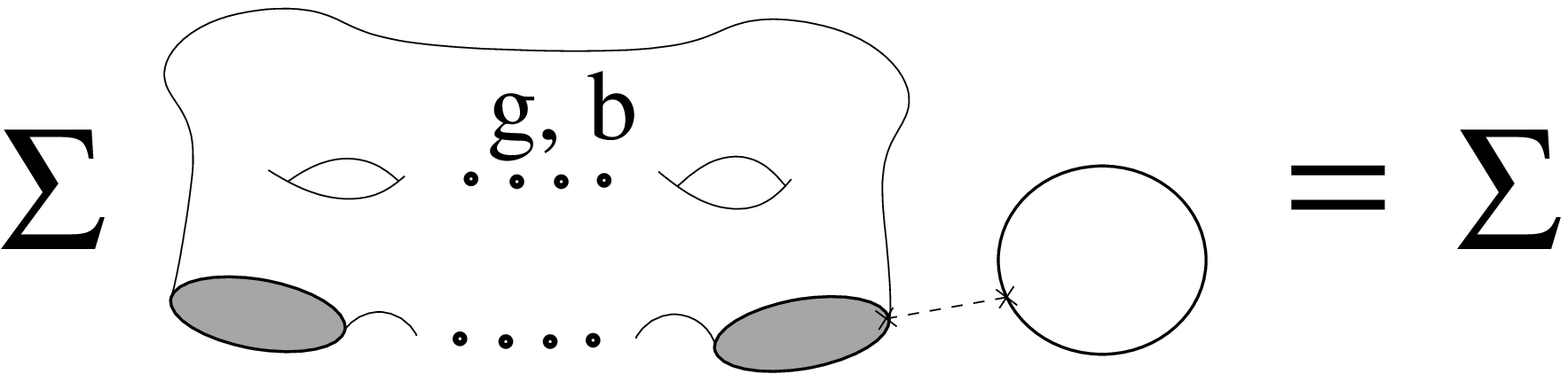}}
    (\textrm{amplitudes with} \ 
    \g > \g_{g,b}) \ .
\eeq
Here, as compared to (\ref{qAinftysum}), we gave up combining the different
levels of genera $g$ into an expansion of the topological string coupling
$g_s$. 
The left-hand side of (\ref{recursive}) comprises all terms from the left-hand
side of (\ref{qAinftysum}) that involve disk amplitudes. Therefore, writing
the quantum $\cA_\infty$-relations in the form (\ref{recursive}) makes apparent
that they provide a sequence of linear systems in
$\cF^{g,b}_{A_1|\ldots|A_b}$, which can be solved recursively, starting from
disk amplitudes.

\section{Quantum master equation?}
\ll{sec:QME}

From string field theory \cite{GZ97OSFT,Kaj01SFT} it is
known that the classical as well as the quantum $\cA_\infty$-structure
have a dual description on a (formal) noncommutative supermanifold. In
our context the latter corresponds to the open string moduli space
(see \cite{Laz05NCgeom} for the precise relation) and we should be able to
recast the quantum $\cA_\infty$-relations (\ref{qAinftysum})
into a quantum master equation on moduli space.

In order to see whether this is indeed true let us introduce for our basis
$\y^{\al\be}_a \in \cH^{\al\be}_{op}$ a dual basis 
$\hat s^{\al\be}_a \in \cH_{op}^d$. The deformation parameters (or
open string moduli) $\hat s^{\al\be}_a$ are 
taken to be associative and graded noncommutative. The latter
requirement accounts for cases where we have Chan--Paton
extensions in the boundary sector \cite{Laz05NCgeom}, \ie, the
deformation parameters are (super)matrices $X_a$. The $\bbZ_2$-degree
of $\hat s^{\al\be}_a$ is the same as the $\bbZ_2$-degree of
$\y^{\al\be}_a$. Let us drop the boundary condition labels again,
understanding that the deformation parameters correspond to edges in
some Quiver diagram associated to the D-brane configuration
\cite{Laz05NCgeom}. 

An element in the ring $A$ of (formal) power series in $\{\hat s_a\}$
is given by $f(\hat s_a) = f_0 + \sum_{m=1}^\infty \frac 1m f_{a_1\ldots a_m} 
\hat s_{a_1} \ldots \hat s_{a_m}$. 
%% Consider furthermore the ring $C^0(A) = A/[A,A]$ of power series
%% $f(s)$ where all coefficients $f_{a_1\ldots a_m}$ are graded cyclic
%% invariant. 
Let us  define left and right partial derivatives
$\lcder_{a}$, $\rcder_{a}: A \rarrow A$ by:
\bea
  \nn
  \lcder_a f(s) &=& (-)^{\td a (\td f + 1)} f(s) \rcder_a = \\
  \nn
  &=&\sum_{m=0}^\infty \sum_{i=1}^{m} 
  (-)^{\td a(\td a_1 + \ldots + \td a_{i-1})}
  f_{a_1\ldots a_{i-1} a a_{i+2} \ldots a_m} 
  \hat s_{a_1} \ldots \hat s_{a_{i-1}} 
  \hat s_{a_{i+1}} \ldots \hat s_{a_m} \ ,
\eea
and the BV operator $\Delta: A \rarrow A$ by:
\[
  \Delta := \rho^{ab} \lcder_a \lcder_b \ .
\]

Consider the formal power series
\beq
  \ll{Qop}
  g_s^{-2} S(g_s,t,\hat s) := \sum_{b=1}^\infty \sum_{m_i = 0}^\infty
    \frac{1}{b! m_1\ldots m_b}
    \cF^b_{A_1|\ldots|A_b}(g_s,t)~ \hat s_{A_1} \ldots \hat s_{A_b}
\eeq 
associated to the all-genus topological string amplitudes
(\ref{allgb}), where we used the abbreviation
$\hat s_{A_i} = \hat s_{a[i]_1}\ldots \hat s_{a[i]_{m_i}}$. Note that
amplitudes with $m_i=0$ are included in the series (\ref{Qop}). It is
understood that $1/{m_i}$ is substituted by $1$ whenever $m_i=0$. From
the $\bbZ_2$-selection rule (\ref{Z2select}) it follows that the
series $S(g_s,t,\hat s)$ has even degree.  

After dressing the quantum $\cA_\infty$-relations (\ref{qAinftysum}) with the
deformation parameters $\hat s_{A_i}$ and summing over all numbers of
boundaries $b$ it follows that the quantum $\cA_\infty$-relations
combine into the quantum master equation 
$\De~e^{-S/g_s^{2}}=0$. Notice however that amplitudes with $m_i=0$
are included in $S(g_s,t,\hat s)$, so that this equation holds only up
to non-stable configurations like in (\ref{non-stable}). We obtain not
quite the quantum master equation, but:
\beq
  \ll{QME}
  \De~ e^{-S/g_s^{2}} = 
  \bigl(- \hat s^\unit \Tr_{\cH^{\al\be}_{op}} (-)^F +~ 
  \textrm{non-stable}~ \bigr)~ 
  e^{-S/g_s^{2}} \ ,
\eeq
which we refer to as the modified quantum master equation.

The converse statement that (\ref{QME}) implies the quantum $\cA_\infty$
relations is not true, because the latter are finer than the modified quantum
master equation. This traces back to definition (\ref{Qop}), from
which we see that $S(g_s,t,\hat s)$ is not a generating function for
the string amplitudes $\cF^{g,b}_{A_1|\ldots|A_b}$. To see this let us
rewrite $S(g_s,t,\hat s)$ in the following way:
\bea
    g_s^{-2} S(g_s,t,\hat s) &=& \sum_{g=0}^\infty \sum_{b=1}^\infty 
    \frac{g_s^{2g+b-2}}{b! m_1\ldots m_b}
    \cF^{g,b}_{A_1|\ldots|A_b}(t)~ \hat s_{A_1} \ldots \hat s_{A_b} 
    \nn \\
    &=& \sum_{-\g=-1}^\infty \sum_{b=1}^{2-\g} 
    \frac{g_s^{-\g}} {b! m_1\ldots m_b}
    \cF^{g,b}_{A_1|\ldots|A_b}(t)~ \hat s_{A_1} \ldots \hat s_{A_b}  
    \nn \\
    &=& \sum_{-\g=-1}^\infty \sum_{b=1}^{2-\g} 
    \frac{g_s^{-\g}} {b! m_1\ldots m_b}
    \cF^{g,b}_{a_1\ldots|\ldots|\ldots a_m}(t)~ 
    \hat s_{a_1} \ldots \hat s_{a_m}  
    \nn \\
    &=& \sum_{-\g=-1}^\infty g_s^{-\g} 
    \cF^{\g}_{a_1\ldots a_m}(t)~ \hat s_{a_1} \ldots \hat s_{a_m}
    \nn \ .
\eea
where $m = \sum_i m_i$ and 
\[\cF^{\g}_{a_1 \ldots a_m}(t)=\sum_{b=1}^{2-\g} 
    \frac{1} {b! m_1\ldots m_b}
    \cF^{g,b}_{a_1\ldots |\ldots|\ldots a_m}(t) \ .
\]
This means that the coefficients in the power series $S(g_s,t,\hat s)$
are sums over 
string amplitudes with the same Euler character and the same boundary
field configuration. However, the genus $g$ as well as the number of
boundaries $b$ vary in this sum. The partitioning of the fields over
the different numbers of boundary components in $\cF^{\g}_{a_1 \ldots a_m}(t)$
must, of course, be consistent with the boundary condition labels.

Therefore, the modified master equation (\ref{QME}) is an equation for
the quantities $\cF^{\g}_{a_1 \ldots a_m}(t)$. The quantum
$\cA_\infty$-relations (\ref{qAinftysum}) are finer in the sense that
they split up with respect to $g$ and $b$.
If we are interested in F-terms for the $4$ dimensional $N=1$
supergravity \cite{BCOV93KS,Vafa00largeN,OV03Cdef,OV03gravCdef} 
or in higher genus open Gromov--Witten invariants, then it is
important to have the more detailed information from (\ref{qAinftysum}).

\section*{Acknowledgments}

I would like to thank Ezra Getzler, Kentaro Hori, Wolfgang Lerche and
Dennis Nemeschansky for enlightening discussions related to the
present work. Special thanks to the KITP in Santa Barbara, as well as the
organizers of the program Mathematical structure of string theory, for
their hospitality and the great research environment during the earlier
stages of this project. This research was supported in part by the
NSERC grant ($\# 458109$).

%% \bibliography{lit1}
%% \bibliographystyle{unsrt}

\end{document}